\documentclass[jmp,preprint,superscriptaddress,showpacs,showkeys]{revtex4-1}

\expandafter\let\csname equation*\endcsname\relax
\expandafter\let\csname endequation*\endcsname\relax
\usepackage{amsmath}
\usepackage{graphicx}
\newcommand{\ud}{\mathop{}\!\mathrm{d}}
\usepackage{xcolor}
\usepackage{amssymb}
\usepackage{hyperref}

\begin{document}

\title{Lorentz Invariant Berry Phase for a Perturbed Relativistic Four Dimensional Harmonic Oscillator}

\author{Y. Bachar}
\affiliation{Dept. of Physics, Bar Ilan Univeristy}
\email{yossi\_dk@yahoo.com}
\author{R. I. Arshansky}
\affiliation{Givat Zorfatit, Etzel 12/14, Jerusalem 9754}
\author{L. P. Horwitz}
\affiliation{School of Physics and Astronomy, Tel-Aviv University}
\affiliation{Dept. of Physics, Bar Ilan Univeristy}
\affiliation{Dept. of Physics, Ariel University in the Shomron}
\author{I. Aharonovich}
\affiliation{Dept. of Physics, Bar Ilan Univeristy}


\begin{abstract}
We show the existence of Lorentz invariant Berry phases generated, 
in the Stueckleberg-Horwitz-Piron manifestly covariant quantum theory (SHP), 
by a perturbed four dimensional harmonic oscillator. 
These phases are associated with a fractional perturbation 
of the azimuthal symmetry of the oscillator. 
They are computed numerically by using time independent perturbation 
theory and the definition of the Berry phase generalized to the framework of 
SHP relativistic quantum theory.
\end{abstract}

\maketitle

A manifestly covariant quantum mechanics was formulated by E. C. G. Stueckelberg \cite{Stu1} in 1941.
He studied this theory for the case of a single particle in an external field. He considered the 
phenomenon of pair annihilation and creation as a manifestation of the development, in each case, 
of a single world line that curves in such a way that one part runs forward and one part runs 
backward in time, and above the turning point the line does not pass at all. 
This configuration was considered by Stueckelberg to represent pair annihilation. 
To describe such a curve, parametrization by the variable $ t $ is ineffective, 
since the trajectory is not single valued. He therefore introduced a parametric description, 
with parameter $ \tau $, along the world line. Hence one branch of the curve is generated 
by motion in the positive sense of $ t $ as a function of increasing $ \tau $, 
and the other branch by motion in the negative sense of $ t $, {\it i.e.}, 
the antiparticle.

The motion, in space-time, of the point generating the world line, 
which we shall call an event (which has the properties of space-time position and energy momentum), 
is governed in the classical case by the Hamilton equations in space-time

\begin{align}
  \label{eq:sctueckelberg_hamiltonian}
  \dfrac{\ud x^{\mu}}
        {\ud \tau   }
  & =
      \dfrac{\partial K}
            {\partial p_{\mu}}
      ,
  &
  \dfrac{\ud p^{\mu}}
        {\ud \tau   }
  & =
      - \dfrac{\partial K}
              {\partial x_{\mu}}
\end{align}

where $ x^{\mu }=(t,\vec{x}) $, $ p^{\mu }=(E,\vec{p}) $ 
[we take $ c=1 $ and $ g_{\mu \nu }=(-1,1,1,1) $] and the 
evolution generator $ K $ is a function of the canonical 
variables $ x_{\mu },p_{\mu } $ which satisfies the 
Poission brackets $\left \{ x_\mu ,p_\nu  \right \}=g_{\mu \nu }$. 
For the special case of free motion,
\begin{align}
    \label{eq:free_hamiltonian}
    K_{0}
    & =
         \dfrac{p^{\mu} p_{\mu}}
               {     2M        }
\end{align}

where $ M $ is an intrinsic parameter assigned to the generic event, and hence
\begin{align}
    \dfrac{\ud x^{\mu}}
          {\ud \tau   }
    & =
         \dfrac{p^{\mu}}
               { M     }
\end{align}

It then follows that
\begin{align}
    \dfrac{\ud \vec{x}}
          {\ud t      }
    & =
        \dfrac{\vec{p}}
              {E      }
\end{align}
consistent with standard relativistic kinematics. We note, however, 
that the mass squared $ m^{2}=-p^{\mu }p_{\mu } $ is a dynamical variable since $\vec{p}$ and 
E are considered to be kinematically independent, 
and therefore it is not taken to be equal to an \emph{a priori} 
given constant. The set of values taken by $ m^{2} $ 
in a particular dynamical context is determined by initial conditions and the dynamical equations.

In the quantum theory, $ \vec{x},t $ (and $ \vec{p},E $) denote operators satisfying the commutation relations (we take $ \hbar=1 $)
\begin{align}
    [x^{\mu}, p^{\nu} ] & = i g^{\mu \nu}
\end{align}

The state of a one-event system is described by a wave function $ \psi_{\tau }(x)\in L^{2}(R^{ 4}) $, a complex Hilbert space with measure $ \ud^{4}x=\ud^{3}x \, \ud t $ satisfying the equation
\begin{align}
i\dfrac{\partial \psi _{\tau }(x)}{\partial \tau }=K\psi _{\tau }(x)
\end{align}

This equation, designed to provide a manifestly covariant description of relativistic phenomena, is similar in form to the non-relativistic Schr\"{o}dinger equation. Although free motion is determined by the operator form of $ K_{0} $ of Eq. (2), {\it i.e.}, the d'Alembertian, which is hyperbolic ($ p_{\mu }p^{\mu  }\equiv-{\partial_{\mu } }{\partial^{\mu }} $ instead of the elliptic operator $ \vec{p}^{2}\equiv -\nabla^{2} $), the same methods may be used for studying Eq. (6) as for the non-relativistic Schrodinger equation.

In 1973 Horwitz and Piron \cite{Piron1973} generalized the Stueckelberg theory by assuming that for the treatment of systems of more than one event (generating world lines of more than one particle), 
there is a single universal $\tau$ which parametrizes the motion of all of the particles of the many body system (we denote this generalized theory by SHP). They assumed the unperturbed evolution generator to be of the form 
\begin{align}
    \label{eq:free_hamiltonian}
    K_{0}
    & =
         \sum_{i=1}^{N}\dfrac{p_i^{\mu} p_{i\mu}}
               {     2M_i        }
\end{align}

There is a class of model systems, for which solutions can be achieved using straightforward methods, which involve only effective action-at-a-distance (direct action) potentials, where the evolution generator is of the form
\begin{align}
   K=\sum_{i=1}^{N}\dfrac{p_{i}^{2}}{2M_{i}}+V(x_{1},x_{2},...,x_{N})
\end{align}

Note that in this case the potential function enters into the dynamical evolution equation as a term added to the generator of the free motion, and therefore corresponds to a space-time coordinate-dependent interaction mass.

In 1989 Horwitz and Arshansky \cite{arshansky:66} demonstrated the existence of bound state solutions for the quantum case by solving the dynamical equation (6) associated with the dynamical evolution operator (8). The two-body potential function, which they chose was of the form $V(\rho)=\frac{k}{\rho}$, where
\begin{align}
\rho =\sqrt{(x_{1}^{\mu }-x_{2}^{\mu })(x_{1\mu }-x_{2\mu })}\equiv \sqrt{(x_{1}-x_{2})^{2}}
\end{align}
and $x_1^{\mu }-x_2^{\mu }$ is spacelike, and $k$ is a coupling constant. For the two body case, the Hamiltonian can be written as a sum of center of mass and relative (reduced motion part) using exactly the same procedure as the nonrelativistic case (the Hamiltonian has the same quadratic form), with potential $V(x)$, where $x= x_1-x_2$ \cite{arshansky:66}.
\newline
Cook solved this problem \cite{Cook1972} with support for $x^\mu$ in the full spacelike region and obtained a spectrum which disagreed with the nonrelativistic Schrodinger spectrum. Zmuidzinas \cite{zm_1966} showed, however that there is no complete orthogonal set in the full spacelike region, but there are complete sets in the two manifolds (defined as in Figure 1), by separating the spacelike region into two parts for which $x^2+y^2-t^2>0$ (The RMS) and $x^2+y^2-t^2<0$.
Then Arshansky and Horwitz defined wavefunctions with support in RMS to obtain a spectrum which agrees with the nonrelativistic Schrodinger solutions. 

\begin{figure}[tp]
    \begin{center}
        \fbox
        {
            \includegraphics[width=10cm]{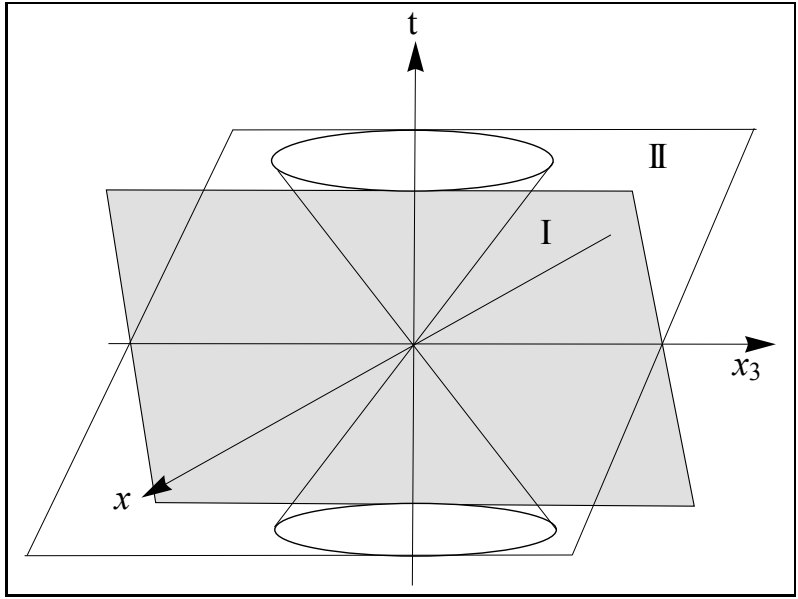}
        }
        \caption
        {
            RMS-Reduced Minkowski Space (I) consists of the space-time points external 
            (in spacelike direction) to the two hyperplanes tangent to the light cone 
            that is oriented along the z axis (the direction must be chosen to define this space). 
            The second subregion (II) consists of the space-time points in the 
            sector interior (timelike direction) to these hyperplanes, but excluding the light cone.
            In this figure the configuration is shown schematically 
            by folding the two space axes x,y together (defining the coordinate $x_{\perp }$). 
            Both subregions are invariant under an $O(2,1)$ subgroup of $O(3,1)$
        } 
        \label{fig:RMS}
    \end{center}
\end{figure}

To describe the RMS we use the definition

\begin{align}
x^{0}=\rho \sin\theta \sinh\beta
  &&  x^{1}=\rho \sin\theta \cos\phi \cosh\beta
  &&  x^{2}=\rho \sin\theta \sin\phi \sinh\beta
  &&  x^{3}=\rho \cos\theta
\end{align}
These variables span the entire RMS. The normalizable solutions of the Stueckelberg-Schrodinger 
equation vanish identically on the boundaries, so there is no tunneling to the second spacelike region (II) or the light cone.

When they applied the method of treating the relativistic quantum two body problem to the case 
of the four dimensional harmonic oscillator, they took the reduced Hamiltonian to be

\begin{align}
\label{eq: the New Hamiltonian}
 \begin{split}
      K=\frac{p_{\mu }p^{\mu }}{2m}+\frac{1}{2}kx_{\mu }x^{\mu }
      &
        =-\frac{1}{2m}\partial_\mu\partial^\mu + \frac{1}{2}mw^2\rho^2
      \\
      &\qquad\
        =\frac{1}{2m}[-\frac{\partial^2 }{\partial \rho ^2}-\frac{3}{\rho }\frac{\partial }{\partial \rho }+
         \frac{\Lambda }{\rho^2 }]+\frac{1}{2}m w^2\rho^2
\end{split}
\end{align}

They obtained the complete set of eigenvectors, 
represented by the $\tau$ independent wave
functions

\begin{align}
\label{eq: the wavefunction}
    \begin{split}
        \psi_{n_alnm}(\phi ,\beta ,\theta ,\rho )
        & =
            \frac{1}{2\pi }e^{i(m+1/2)\phi }
            \times
            \sqrt{n}
            \sqrt{\Gamma (1+m+n)/\Gamma (1+m-n)}
            \times
        \\
        & \qquad \times
            (1-\tanh^2\beta )^\frac{1}{4}
            P^{-n}_m(\tanh\beta )
            \times
            (1-\cos^2\theta )^{-\frac{1}{4}}
            \times
        \\
        & \qquad \times
            P^n_l(\cos\theta )
            \frac{1}{\sqrt{\rho }}
            (\frac{mw\rho ^2}{\hbar})^{l/2}
            e^{-\frac{mw\rho ^2}{2\hbar}}
            L^{l+1/2}_{n_a}(\frac{mw\rho ^2}{\hbar})
    \end{split}
\end{align}
We now study the effect of an adiabatic perturbation on the Hamiltonian \eqref{eq: the New Hamiltonian} on the wavefunction \eqref{eq: the wavefunction}.
\newline
The idea of the geometric phase proposed by Berry \cite {Berry1984},\cite {QM2} in 1984 asserts that under adiabatic processes the wave function of a system picks up a phase factor that can be found in the nonrelativistic case by the integral
\begin{align}
\gamma _n (t) \equiv i\int_{0}^{t} \left \langle \psi _n(t') \middle| \frac{\partial\psi _n(t') }{\partial t}\right \rangle dt'
\end{align}
where $\psi_n$ is the nth eigenvector of the slowly changing Hamiltonian (for which one can prove the adiabatic theorem), $t$ is the time between the initial and final values of the time dependent parameters of the Hamiltonian and $\frac{\partial \psi_n}{\partial t}$ corresponds to the change in the wavefunction under time variation of the parameters. To construct a manifestly covariant form of (13), we have to replace the wavefunction by solutions of Eq.(6), and the variable $t$ by the variable $\tau$ which is the evolution time according to SHP quantum mechanics. Suppose there are N $\tau$ dependent parameters: $R_1(\tau),R_2(\tau),.....R_N(\tau) $ in the Hamiltonian of a given problem; then
\begin{align}
\begin{split}
     \frac{\partial \psi _n}{\partial \tau}=\frac{\partial \psi _n}{\partial R_1}\frac{dR_1}{d\tau}+\frac{\partial \psi _n}{\partial R_2}\frac{dR_2}{d\tau}+.......+\frac{\partial \psi _n}{\partial R_N}\frac{dR_N}{d\tau}=(\nabla_{\vec{R}}\psi _n)\cdot \frac{d\vec{R}}{d\tau}
\end{split}
\end{align}
where
\begin{align}
 \vec{R}=(R_1,R_2,....,R_N).
\end{align}
We then have
\begin{align}
\gamma _n(\tau)=i\int_{\vec{R_i}}^{\vec{R_f}} \left \langle\psi _n\mid \nabla_{\vec{R}}\psi _n \right \rangle\cdot d\vec{R}
\end{align}
Now, if the Hamiltonian returns to its original form after a time $\tau = T$, then the geometric phase is
\begin{align}
\label{eq:Berry phase}
\gamma _n(T)=i\oint  \left \langle\psi _n\mid \nabla_{\vec{R}}\psi _n \right \rangle\cdot d\vec{R}
\end{align}
We wish to demonstrate the realization of an {\it invariant} Berry phase in the example of the four dimensional oscillator. For that purpose we add a perturbation to the Hamiltonian (11) which breaks the hyperangular symmetry of the Hamiltonian.
Since the complex valued matrix elements necessary to develop a dynamical phase arise in this example from the $\phi$ dependance, one must perturb the azimuthal symmetry with a fractional coefficient, as we shall see below. As a simple example we take the perturbed Hamiltonian to be
\begin{align}
\label{eq:The New Hamiltonian}
 \begin{split}
     K & =  \frac{1}{2m}
             \Big[
                -\frac{\partial^2 }
                      {\partial \rho ^2}
                -
                \frac{3}{\rho }
                \frac{\partial }{\partial \rho }
                +
                \frac{\Lambda }{\rho^2 }
             \Big]
             +
             \frac{1}{2}m w^2\rho^2
             +
         \\
         & \qquad
                +
             \varepsilon_1\rho^2 \sin^2 \theta \cos^2 ((2/3)\phi) \cosh^2\beta
                +
         \\
         & \qquad
                +
             \varepsilon_2\rho^2 \sin^2 \theta \sin^2((2/3)\phi) \cosh^2\beta
                +
         \\
         & \qquad
                +
             \varepsilon_3\rho^2\cos^2\theta
                -
             \varepsilon_0\rho^2\sin^2\theta \sinh^2\beta
\end{split}
\end{align}
where $\varepsilon _0,\varepsilon _1,\varepsilon_2, \varepsilon_3 $ are the small parameters of the perturbation.
\newline
Now using degenerate time independent perturbation theory, we calculate the first order correction to the wave function (12).
\newline

The new wave function produced by the Hamiltonian (18) is equal to a linear combination of the wave function (12) and its first order correction, which is given by
\begin{align}
 \begin{split}
 \psi _{n_a'l'n'm'}^{(1)}=\sum _{n_a,l\neq n_a',l'}\frac{  \left \langle  \psi _{n_alnm}\mid V\mid \psi _{n_a'l'n'm'}\right \rangle} {K'_a-K_a}\psi _{n_alnm}
\end{split}
\end{align}
where V is the perturbation given in (18), and $K_a$ is the eigenvalue of the unperturbed Hamiltonian (11),
\begin{align}
 \begin{split}
K_a=\hbar\omega \left (l+2n_a+\frac{3}{2} \right )
\end{split}
\end{align}
Now suppose $\varepsilon _0$ and $\varepsilon _3$ are equal to zero so that Eq.(19) becomes
\begin{align}
 \label{eq:first_order_correction_explicit}
 \begin{split}
  \psi _{n'_al'n'm'}^{(1)}=\varepsilon _1\sum _{n_a,l\neq n'_a,l'}\frac{\left \langle\psi _{n_alnm}\mid \rho ^{2}\sin^{2}\theta \cos^{2}((2/3)\phi ) \cosh^{2}\beta \mid \psi _{n'_al'n'm'}\right \rangle}{[l'-l+2(n_a'-n_a)]\hbar\omega } \psi _{n_alnm}
  \\
  \qquad
   +\varepsilon _2\sum _{n_a,l\neq n_a',l'}\frac{\left \langle\psi _{n_alnm}\mid \rho ^{2}\sin^{2}\theta \sin^{2}((2/3)\phi) \cosh^{2}\beta \mid \psi _{n_a'l'n'm'}\right \rangle}{[l'-l+2(n_a'-n_a)]\hbar\omega} \psi _{n_alnm}
\end{split}
\end{align}
or
\begin{align}
 \begin{split}
\psi_{n'_al'n'm'}^{(1)}=\varepsilon _1\psi'_{n'_al'n'm'}+\varepsilon _2\psi''_{n'_al'n'm'}
\end{split}
\end{align}
The simplest way to demonstrate the existence of a geometric phase is by letting each one of the indices 
$(n_a,n'_a,l,l',n,n',m,m')$ take the value 2 or 3. 
In that way our perturbation can be represented as a $16\times16$ matrix. 
The Hilbert space of the problem is a 16 dimensional space composed of four 
4 dimensional subspaces, 
each subspace corresponding to one of the four energy eigenvalues.
The next table summarizes all the possible permutations 
(note that some of the entries in this table, those for 
 which $n >l$, vanish identically and therefore do not 
 appear in any calculation. We list them for completeness among the indices.)
\newline
  \begin{tabular}{|l|l|l|l|l|l|c}
        \hline
                $n_{a}$     &     $l$     &    $n$     &     $m$     &     $i$     &     $K_{ai}$ ($\hbar\omega$)
        \\
         \hline
                    2       &      2      &      2     &       2     &       1     &       7.5
        \\
         \hline
                    2       &      2      &      2     &       3     &       2     &
        \\
         \hline
                    2       &      2      &      3     &       2     &       3     &
        \\
         \hline
                    2       &      2      &      3     &       3     &       4     &
        \\
         \hline
                    2       &      3      &      2     &       2     &       5     &       8.5
        \\
         \hline
                    2       &      3      &      2     &       3     &       6     &
        \\
         \hline
                    2       &      3      &      3     &       2     &       7     &
        \\
         \hline
                    2       &      3      &      3     &       3     &       8     &
        \\
         \hline
                    3       &      2      &      2     &       2     &       9     &       9.5
        \\
         \hline
                    3       &      2      &      2     &       3     &       10    &
        \\
         \hline
                    3       &      2      &      3     &       2     &       11    &
        \\
         \hline
                    3       &      2      &      3     &       3     &       12    &
        \\
         \hline
                    3       &      3      &      2     &       2     &       13    &      10.5
        \\
         \hline
                    3       &      3      &      2     &       3     &       14    &
        \\
         \hline
                    3       &      3      &      3     &       2     &       15    &
        \\
         \hline
                    3       &      3      &      3     &       3     &       16    &
        \\
         \hline
  \end{tabular}
\newline
\newline
Let $D_{1}$ be the subspace of all the eigenfunctions that correspond to the eigenvalue $K_{a1}$, so we can write
\newline
$D_{1}=span\left \{ \psi_{1}, \psi_{2}, \psi_{3}, \psi_{4} \right \}$
\newline
and similarly
\newline
$D_{2}=span\left \{ \psi_{5}, \psi_{6}, \psi_{7}, \psi_{8} \right \}$
\newline
$D_{3}=span\left \{ \psi_{9}, \psi_{10}, \psi_{11}, \psi_{12} \right \}$
\newline
$D_{4}=span\left \{ \psi_{13}, \psi_{14}, \psi_{15}, \psi_{16} \right \}$
\newline
\newline
For example, let us find the first order correction of $\psi _{1}\in D_{1}$; we chose $\psi _{1}$ because this is one of the nonzero ground states (we will see later why it is nonzero) of the oscillator (along with $\psi _{2}$).
\newline

Substituting the matrix elements of V' and V'' which we shall compute numerically later, where 
\begin{align}
V'=\rho ^{2}\sin^{2}\theta \cos^{2}((2/3)\phi ) \cosh^{2}\beta
\end{align}
and
\begin{align}
V''=\rho ^{2}\sin^{2}\theta \sin^{2}((2/3)\phi) \cosh^{2}\beta
\end{align}
while
$\hbar=6.626\cdot10^{-34}J\cdot s$, $m=9.109\cdot10^{-31}kg$ (electron mass), and $\omega=240.4MHz$.
\newline
the value of $\omega$ is determined by the wish to obtain an order one value for the Berry phase (up to the scale of the perturbation squared).
\newline
According to Eq. 
\eqref{eq:first_order_correction_explicit},
\begin{align}
\label{eq:first order correction example}
\begin{split}
\psi _{1}^{(1)} &=
               \varepsilon _{1} \Big[
                \frac{\left \langle\psi _{5}
                \mid V'\mid \psi _{1}
                \right \rangle}{K_{a1}-K_{a5}}\psi _{5}
                +
                \frac{\left \langle\psi _{6}
                \mid V'\mid \psi _{1}
                \right \rangle}{K_{a1}-K_{a6}}\psi _{6}
                +
                \frac{\left \langle\psi _{7}
                \mid V'\mid \psi _{1}
                \right \rangle}{K_{a1}-K_{a7}}\psi _{7}
    \\
    \\
    & \qquad     + \frac{\left \langle\psi _{8}
                \mid V'\mid \psi _{1}
                \right \rangle}{K_{a1}-K_{a8}}\psi _{8}
                +
                  \frac{\left \langle\psi _{9}
                \mid V'\mid \psi _{1}
                \right \rangle}{K_{a1}-K_{a9}}\psi _{9}
                +
                  \frac{\left \langle\psi _{10}
                \mid V'\mid \psi _{1}
                \right \rangle}{K_{a1}-K_{a10}}\psi _{10}
     \\
     \\
    & \qquad     +\frac{\left \langle\psi _{11}
                \mid V'\mid \psi _{1}
                \right \rangle}{K_{a1}-K_{a11}}\psi _{11}
                +
                  \frac{\left \langle\psi _{12}
                \mid V'\mid \psi _{1}
                \right \rangle}{K_{a1}-K_{a12}}\psi _{12}
                +
                  \frac{\left \langle\psi _{13}
                \mid V'\mid \psi _{1}
                \right \rangle}{K_{a1}-K_{a13}}\psi _{13}
      \\
      \\
    & \qquad    +  \frac{\left \langle\psi _{14}
                \mid V'\mid \psi _{1}
                \right \rangle}{K_{a1}-K_{a14}}\psi _{14}
                +
                  \frac{\left \langle\psi _{15}
                \mid V'\mid \psi _{1}
                \right \rangle}{K_{a1}-K_{a15}}\psi _{15}
                +
                  \frac{\left \langle\psi _{16}
                \mid V'\mid \psi _{1}
                \right \rangle}{K_{a1}-K_{a16}}\psi _{16}
                \Big]
       \\
       \\
   &
              +
                \varepsilon _{2} \Big[
                \frac{\left \langle\psi _{5}
                \mid V''\mid \psi _{1}
                \right \rangle}{K_{a1}-K_{a5}}\psi _{5}
                +
                \frac{\left \langle\psi _{6}
                \mid V''\mid \psi _{1}
                \right \rangle}{K_{a1}-K_{a6}}\psi _{6}
                +
                \frac{\left \langle\psi _{7}
                \mid V''\mid \psi _{1}
                \right \rangle}{K_{a1}-K_{a7}}\psi _{7}
    \\
    \\
    & \qquad    +\frac{\left \langle\psi _{8}
                \mid V''\mid \psi _{1}
                \right \rangle}{K_{a1}-K_{a8}}\psi _{8}
                +
                  \frac{\left \langle\psi _{9}
                \mid V''\mid \psi _{1}
                \right \rangle}{K_{a1}-K_{a9}}\psi _{9}
                +
                  \frac{\left \langle\psi _{10}
                \mid V''\mid \psi _{1}
                \right \rangle}{K_{a1}-K_{a10}}\psi _{10}
     \\
     \\
    & \qquad    + \frac{\left \langle\psi _{11}
                \mid V''\mid \psi _{1}
                \right \rangle}{K_{a1}-K_{a11}}\psi _{11}
                +
                  \frac{\left \langle\psi _{12}
                \mid V''\mid \psi _{1}
                \right \rangle}{K_{a1}-K_{a12}}\psi _{12}
                +
                  \frac{\left \langle\psi _{13}
                \mid V''\mid \psi _{1}
                \right \rangle}{K_{a1}-K_{a13}}\psi _{13}
      \\
      \\
    & \qquad    +  \frac{\left \langle\psi _{14}
                \mid V''\mid \psi _{1}
                \right \rangle}{K_{a1}-K_{a14}}\psi _{14}
                +
                  \frac{\left \langle\psi _{15}
                \mid V''\mid \psi _{1}
                \right \rangle}{K_{a1}-K_{a15}}\psi _{15}
                +
                  \frac{\left \langle\psi _{16}
                \mid V''\mid \psi _{1}
                \right \rangle}{K_{a1}-K_{a16}}\psi _{16}
                \Big]
 \end{split}
\end{align}
Hence, the new wave function is a linear combination of the original eigenfunction and its first order correction
\begin{align}
\Psi_{1}=\psi _{1}+\psi _{1}^{(1)}=\psi _{1}+\varepsilon _{1}\psi'_{1}+\varepsilon _{2}\psi ''_{1}
\end{align}
(where we replace the indices $n'_a,l',n',m'$ by the index $j$, in the same way we replace the indices $n_a,l,n,m$ by the index $i$, according to the table in page 6)
\newline
We wish to calculate the Berry phase produced by the new wave function. For that purpose we first evaluate the formula of the Berry phase for the case of the perturbed four dimensional harmonic oscillator and then we will apply this formula to compute the particular Berry phase produced by the wave function $\psi _{1}$
\newline
We take the parameter space to be $\vec{R}=\vec{R}(\varepsilon _{1},\varepsilon _{2})$ where $\varepsilon _{1},\varepsilon _{2}$ are the coupling parameters of the perturbation.
Hence
\begin{align}
\label{eq:inner product}
\begin{split}
\left \langle \Psi_{j} \mid \nabla_{\vec{R}}\Psi_{j} \right \rangle &=
\left \langle \psi _{j}+\varepsilon _{1}\psi'_{j}+\varepsilon _{2}\psi ''_{j}\mid \psi'_{j}\hat{\varepsilon _{1}}
+ \psi ''_{j}\hat{\varepsilon _{2}}\right \rangle
\\
&  =\left \langle \psi _{j}\mid \psi '_{j} \right \rangle\hat{\varepsilon _{1}}
         +\left \langle \psi _{j}\mid \psi ''_{j} \right \rangle\hat{\varepsilon _{2}}
\\
&  +\varepsilon _{1}\left \langle \psi' _{j}\mid \psi '_{j} \right \rangle\hat{\varepsilon _{1}}
         +\varepsilon _{1}\left \langle \psi' _{j}\mid \psi ''_{j} \right \rangle\hat{\varepsilon _{2}}
\\
&  +\varepsilon _{2}\left \langle \psi'' _{j}\mid \psi '_{j} \right \rangle\hat{\varepsilon _{1}}
         +\varepsilon _{2}\left \langle \psi'' _{j}\mid \psi ''_{j} \right \rangle\hat{\varepsilon _{2}}
\end{split}
\end{align}
where $\hat{\varepsilon _{1}}$ and $\hat{\varepsilon _{2}}$ are unit vectors in the parameter space.
\newline
\newline
Now, let us define
\begin{align}
\psi_{j} '=\sum_{i\neq j}a_{i}\psi _{i} \qquad \psi_{j} ''=\sum_{i\neq j}b_{i}\psi _{i}
\end{align}
Hence, from the orthonormality of these eigenfunctions,
\begin{align}
\left \langle \psi _{j}\mid \psi '_{j} \right \rangle=\left \langle \psi _{j}\middle|  \sum_{i\neq j}a_{i}\psi _{i} \right \rangle=\sum_{i\neq j}a_{i}\left \langle \psi _{j}\mid  \psi _{i}\right \rangle=0
\end{align}
and similarly
\begin{align}
\left \langle \psi _{j}\mid \psi ''_{j} \right \rangle=0
\end{align}
We also have that
\begin{align}
\left \langle \psi '_{j}\mid \psi '_{j} \right \rangle=\left \langle  \sum_{i\neq j}a_{i}\psi _{i}\middle| \sum_{k\neq j}a_{k}\psi _{k}\right \rangle=\sum_{i\neq j}a^{*}_{i}\sum_{k\neq j}a_{k}\left \langle \psi _{i}\mid \psi _{k} \right \rangle=\sum_{i\neq j}a^{*}_{i}\sum_{k\neq j}a_{k}\delta _{ik}=\sum_{i\neq j}\left | a_{i} \right |^{2}
\end{align}
and similarly
\begin{align}
\left \langle \psi ''_{j}\mid \psi ''_{j} \right \rangle=\sum_{i\neq j}\left | b_{i} \right |^{2}
\end{align}
\begin{align}
\label{important}
\left \langle \psi '_{j}\mid \psi ''_{j} \right \rangle=\left \langle \sum_{i\neq j}a_{i}\psi_{i}\middle| \sum_{k\neq j}b_{k}\psi_{k} \right \rangle=\sum_{i\neq j}a^{*}_{i}\sum_{k\neq j}b_{k}\left \langle \psi_{i}\mid \psi_{k}  \right \rangle=\sum_{i\neq j}a^{*}_{i}\sum_{k\neq j}b_{k}\delta_{ik}=\sum_{i\neq j}a^{*}_{i}b_{i}
\end{align}
Substituting these results into Eq.\eqref{eq:inner product}, we get
\begin{align}
\label{eq:inner product 2}
\begin{split}
\left \langle \Psi_{j} \mid \nabla_{\vec{R}}\Psi_{j} \right \rangle=\left [  \varepsilon _{1}\sum_{i\neq j}\left | a_{i} \right |^{2}+\varepsilon _{2}\sum_{i\neq j}a_{i}b^{*}_{i}\right ]\hat{\varepsilon _{1}}+\left [\varepsilon _{1}\sum_{i\neq j}a^{*}_{i}b_{i}+\varepsilon _{2}\sum_{i\neq j}\left | b_{i} \right |^{2}\right ]\hat{\varepsilon _{2}}
\end{split}
\end{align}
Now, since $\vec{R}=\vec{R}(\varepsilon _{1},\varepsilon _{2})$ then
\begin{align}
d\vec{R}=d\varepsilon _{1}\hat{\varepsilon _{1}}+d\varepsilon _{2}\hat{\varepsilon _{2}}
\end{align}
Hence
\begin{align}
\label{eq:inner product 3}
\left \langle \Psi_{j} \mid \nabla_{\vec{R}}\Psi_{j} \right \rangle\cdot d\vec{R}=\left [  \varepsilon _{1}\sum_{i\neq j}\left | a_{i} \right |^{2}+\varepsilon _{2}\sum_{i\neq j}a_{i}b^{*}_{i}\right ]d\varepsilon _{1}+\left [\varepsilon _{1}\sum_{i\neq j}a^{*}_{i}b_{i}+\varepsilon _{2}\sum_{i\neq j}\left | b_{i} \right |^{2}\right ]d\varepsilon _{2}
\end{align}
Now, let us assume that the adiabatic change of the parameters $\varepsilon _{1}$ and $\varepsilon _{2}$ follow a circle (with small radius $\it{r}$) in the parameter space, so we take $\varepsilon _{1}=r\cos\alpha$ and $\varepsilon _{2}=r\sin\alpha$.
From Eq.\eqref{eq:inner product} and Eq.\eqref{eq:Berry phase}, we obtain
\begin{align}
\begin{split}
\gamma _{j} =
         ir^2\Big[-\sum_{i\neq j}\left | a_{i} \right |^{2}
          \int_{0}^{2\pi }\cos\alpha\sin\alpha d\alpha
        - \sum_{i\neq j}a_{i}b^{*}_{i}\int_{0}^{2\pi }\sin^{2}\alpha d\alpha
       \\
         +\sum_{i\neq j}a^{*}_{i}b_{i}
          \int_{0}^{2\pi }\cos^{2}\alpha d\alpha
         +\sum_{i\neq j}\left | b_{i} \right |^{2}
          \int_{0}^{2\pi }\sin\alpha \cos\alpha d\alpha \Big]
\end{split}
\end{align}
Noting that the first and fourth integrals in this expression are equal to zero, and that the second and third integrals are equal to $\pi$, we get
\begin{align}
\gamma _{j}=i\pi r^2 \sum_{i\neq j}(a^{*}_{i}b_{i}-a_{i}b^{*}_{i})=-2\pi r^2 \sum_{i\neq j}Im(a^{*}_{i}b_{i}) =-2\pi r^2 Im\left \{ \sum_{i\neq j}a^{*}_{i}b_{i} \right \}
\end{align}
or, in terms of the perturbed wavefunction (by \eqref{important}) we find that
\begin{align}
\label{eq:gamma}
\gamma _{j}=-2\pi r^2 Im\left \langle \psi '_{j}\mid \psi ''_{j} \right \rangle
\end{align}
This result can obtained directly from (27) by noting that $dR = -r\sin\alpha d\alpha \hat{\varepsilon _1}+ r\cos\alpha d\alpha \hat{\varepsilon _2}$; therefore only terms in  $\varepsilon_1 d\varepsilon_2$ and  $\varepsilon_2 d\varepsilon_1$ can contribute to the integral. The result is exactly as given in (39). We have given the orthogonal expansion method as well since it seems instructive to do so.
\newline
Hence, the covariant Berry phase, produced by the wave function $\psi_1$ (for example) of the perturbed harmonic oscillator, is
\begin{align}
\gamma _{1}/r^2=-2\pi Im\left \langle \psi '_{1}\mid \psi ''_{1} \right \rangle=1.057
\end{align}
\newline

\begin{table}
 \begin{center}
  \begin{tabular}{|c|c|c|}
        \hline
                   $j$     &  $\omega$     &    $\gamma_{j}/r^2$
        \\
         \hline
                    1      &   240.4MHz    &      1.057
        \\
         \hline
                    2      &   89.6MHZ     &      6.429
        \\
         \hline
                    5      &   240.4MHz    &      2.470
        \\
         \hline
                    6      &   240.4MHz    &      2.095
        \\
         \hline
                    8      &   334.02MHz   &      2.840
        \\
         \hline
                    9      &   240.4MHz    &      7.905
        \\
         \hline
                    10     &   89.6MHz     &      6.429
        \\
         \hline
                    13     &   240.4mHz    &      2.470
        \\
         \hline
                    14     &   240.4MHz    &      2.095
        \\
         \hline
                    16     &   334.02MHz   &      2.840
        \\
         \hline
\end{tabular}
     \caption{Table of angular frequencies and geometric phases}
     \label{tab:angular_freqency_table}
\end{center}
\end{table}

Formula \eqref{eq:gamma} contains a scalar product on a given orientation of the RMS, 
say, $n^\mu=(0,0,0,1)$, labeling the defining representation according to Figure 1. 
A Lorentz transformation of the system induces a transformation of coordinates 
in $O(2,1)$ of the RMS (the "little group" of the induced representation \cite{arshansky:380}) 
and a reorientation along the orbit. The perturbation in Eq. \eqref{eq:The New Hamiltonian} does not contain $n^\mu$ explicitly, 
and therefore the perturbed wave functions depend on $n^\mu$ only through the coordinatization of the RMS.
\newline
In the scalar product \eqref{eq:gamma}, one may make a change of variables to 
compensate for the induced $O(2,1)$ transformation on the variables of the RMS. 
The z component, oriented in any frame along the $n^\mu$ axis, is invariant, 
and therefore the scalar product defines a set of {\it invariant} phases $\gamma _{n}$

We have demonstrated the existence of invariant Berry phases in a relativistically 
covariant four dimensional harmonic oscillator. As we showed, these phases must 
be associated with fractional angle perturbation of the azimuthal symmetry of the oscillator.
As can be expected, the Berry phase is zero for each eigenstate with 
$l< n$ (so $P_{l}^{n}(\cos\theta)=0$ and hence, 
the oscillator's eigenfunction is identically zero) or 
$m< n$ (so $P_{m}^{n}(\tanh\beta)=0$ and hence, 
the first order correction to the oscillator's eigenfunction is equal to zero), 
and is nonzero for all the other eigenstates (where $l\geqslant n$ and $m\geqslant n$), We can see their
specific values in Table  \ref{tab:angular_freqency_table}.
\newline
\newline
We have shown, furthermore, that, since the perturbing potential does not depend explicitly 
on the orientation of the RMS according to $n^\mu$, the Berry phases we have computed are {\it Lorentz invariant}.

\vspace{6mm}


\end{document}